\documentstyle[preprint,aps,prb]{revtex}
\begin{document}
\draft
\title{
The Schr\"odinger formulation of the 
Feynman path centroid density         
}
\author{Rafael Ram\'{\i}rez and Telesforo L\'opez-Ciudad} 
\address{Instituto de Ciencia de Materiales,
         Consejo Superior de Investigaciones Cient\'{\i}ficas
         (C.S.I.C.),
         Cantoblanco, 28049 Madrid, Spain   }
\date{\today}
\maketitle
\begin{abstract}
We present an analysis of the Feynman path centroid density that
provides new insight into the correspondence between the
path integral and the Schr\"odinger formulations of statistical mechanics.
The path centroid density is a central concept for      
several approximations 
(centroid molecular dynamics, quantum transition state
theory, and pure quantum self-consistent harmonic approximation) that are used
in path integral studies of thermodynamic and dynamical 
properties of quantum particles.
The centroid density is related to   
the quasi-static response of the equilibrium system to an external force.
The path centroid dispersion is the canonical correlation of the
position operator, that measures the linear 
change in the mean position of a quantum particle upon the application of
a constant external force. 
At low temperatures, this quantity 
provides an approximation to the 
excitation energy of the quantum system. 
In the zero temperature limit, the particle's probability density
obtained by fixed centroid path integrals corresponds to the
probability density of {\it minimum energy wave packets},
whose average energy define the Feynman effective 
classical potential.

\end{abstract}
\section{INTRODUCTION}

The thermodynamic properties of many quantum
systems
have been studied within the path integral formulation of statistical
mechanics.
\cite{ceperley95}
An essential aspect of this formulation is the mapping
of the quantum system onto a classical model, 
whose equilibrium properties can be derived
by classical molecular dynamics or Monte Carlo simulation techniques  
with arbitrarily accuracy.
However,
the efficient calculation of equilibrium dynamical properties
of quantum many-body systems
by this formulation,
i.e., time dependent correlation functions,
still remains as a challenging unsolved problem.
\cite{gubernatis96}

Several applications of the path integral formulation 
have been developed around a quantity introduced by Feynman and Hibbs: 
\cite{feynman65},
the {\it effective classical potential} (ECP). 
This concept allows the formulation of a variational principle
for quantum systems in thermodynamic equilibrium,
whose importance can be appreciated by
quoting the conclusions of the 
Feynman and Hibbs' book: "[This variational principle is] the
only example of a result obtained with path integrals which cannot
be obtained in simple manner by more conventional methods". 
\cite{feynman65bis}
The original formulation of this principle used
a free particle as a reference system, a fact that limits the
range of validity of this approximation to temperatures 
where the system behaves nearly in a classical way.
An essential improvement of this variational theory was
formulated by Giachetti and Tognetti \cite{giachetti86}, and,
independently, by Feynman and Kleinert \cite{feynman86}, using
a harmonic oscillator as reference system. With this improvement
the equilibrium properties of quantum anharmonic systems can be
realistically approximated even in the low temperature limit.
\cite{acocella95} The name "pure quantum self-consistent
harmonic approximation" has been recently coined 
for the application of the 
variational principle at this level of approximation. \cite{cuccoli95}

Further applications of the ECP concept are
related to kinetic and dynamical properties of quantum 
systems in thermodynamic equilibrium. Gillan 
formulated the basis of the {\it quantum transition state theory} (QTST),
a kinetic approximation to calculate rate constants of thermal
activated processes, that may be applied to anharmonic many-body systems
at temperatures where quantum effects are
important. \cite{gillan87,gillan90}  
This approximation has been used
in condensed matter and chemical physics investigations, e.g. 
in recent studies of reorientation rates of hydrogen around a boron atom
in doped silicon, \cite{noya97}, or proton transfer
reactions \cite{lobaugh94}. The most important quantity
in QTST is the so called 
{\it centroid density}, a function that carries exactly the 
same information as the ECP: the centroid density is an exponential
function of the ECP. An interesting dynamical approximation, called
{\it centroid molecular dynamics} (CMD), was
formulated by Cao and Voth \cite{cao94} to calculate
real time correlation functions of quantum
particles at finite temperatures.
The ECP is an essential quantity for
this approximation as the dynamical properties 
are derived from trajectories generated by classical
equations of motion of particles moving in the ECP. 
A justification of the CMD approach has been given by showing that 
the centroid position correlation function agrees to second
order, in a Taylor expansion in time, with
the Kubo transformed position correlation function. \cite{cao94}
The latter correlation function is related to the dynamical response
of the system to an external force.
Unfortunately, the limits and capability of CMD for many-body
systems at temperatures where quantum effects and anharmonicity 
are relevant have not been at present fully established.
For recent applications  
of this dynamical approach see Refs. \cite{voth96} and \cite{klein98}.

The centroid density or, equivalently, the
ECP, are then important concepts in the
theory of path integrals. The motivation of the present work is
based on a simple idea: if the centroid density
is the basis of several interesting physical applications
within the path integral formulation (i.e., a variational
principle, and the QTST and CMD approximations), then
the definition of this quantity within the
Schr\"odinger formulation may lead to new physical insight 
into these approximations. Our goal is to present the correspondence
between path integral concepts, such as the centroid density or the ECP,
and the Schr\"odinger formulation.
In a recent contribution \cite{ramirez98}
we have analyzed the centroid density showing that
is related to the quasi-static response 
of the quantum system to a constant external force. Some physical implications
of this analysis have been explored in the zero temperature limit, e.g.,
the ECP corresponds
to the mean energy of {\it minimum energy  
wave packets} (MEWP, to be defined below) 
and CMD is an approximate dynamics
based on these wave packets.
In this work we present a full account of these findings,
with focus on static equilibrium properties. 

This paper is organized as follows. The theoretical part
is presented in Sec. \ref{2}, which is divided
into several subsections.  
Sections \ref{2A} and \ref{2B} review the Schr\"odinger 
and path integral formulation 
of the equilibrium density matrix. The influence of a 
constant external force on the Euclidean action of the system
is treated in Sec. \ref{2C}. Important path integral
quantities, as the centroid density and the static-force
response (SFR) density matrix,   
are introduced in Sec. \ref{2D}. In Sec. \ref{2E}
the centroid density and the SFR density matrix are derived
in the Schr\"odinger formulation by means of their moment
generating functions.
Sec. \ref{3} presents some relevant physical consequences of the 
theoretical part. It includes two subsections, Sec. \ref{3A}
deals with the static isothermal susceptibility
at finite temperatures, and Sec \ref{3B}
treats the zero temperature limit of the ECP and of the SFR density matrix.
The concluding remarks are given in Sec. \ref{4}.

\section{Theory}
\label{2}

\subsection{Schr\"odinger formulation of the density matrix}
\label{2A}

We consider the simple case 
of a quantum particle of mass $m$ 
having bound states in a one-dimensional potential
$V(x)$. 
The extension to the many-particle case, for distinguishable
particles, is presented in Appendix \ref{app1}.
For the subsequent analysis of the Feynman path centroid density, we
study the static response of the particle
to a constant external force, $f$. Therefore, we consider the Hamiltonian
of the particle, $\hat{H}(f)$, as a function of the external force:

\begin{equation}
\hat{H}(f) = \hat{H} - f \hat{x}   \; ,
\end{equation}

where $\hat{x}$ is the position operator and
$\hat{H} \equiv \hat{H}(0)$ is the 
Hamiltonian operator in the absence of the force:

\begin{equation}
\hat{H} = \frac{\hat{p}^2}{2m} + V(\hat{x}) \; ,
\end{equation}

with $\hat{p}$ being the momentum operator.
The following convention is used throughout the text: operators 
(or functions) that share the same name, as $\hat{H}(f)$ and $\hat{H}$,
but differ either by the presence or
number of arguments, represent different operators (or functions).
The eigenfunctions, $|\psi_n(f) \rangle $, of the Hamiltonian,
$\hat{H}(f)$,  satisfy the
Schr\"odinger equation:

\begin{equation}
\hat{H}(f) \,\, |\psi_n(f) \rangle = E_n(f) \,\, |\psi_n(f) \rangle \; ,
\end{equation}

where $E_n(f)$ are the corresponding eigenvalues. In the position
representation, the eigenfunctions are expressed as:

\begin{equation}
\psi_n(x;f) \equiv \langle x | \psi_n(f) \rangle  \; , 
 \text{with} \; n= 0,1,2,...  \; .
\end{equation}

If we consider a canonical ensemble of independent 
particles in thermal equilibrium
at temperature $T$, the statistical state of the system in the presence
of the external force, $f$, is described
in terms of the unnormalised density operator:

\begin{equation}
\hat{\rho}(f) = e^{-\beta \hat{H}(f)}  \; ,
\end{equation}

where $\beta$ is the inverse temperature $(k_B T)^{-1}$, with
$k_B$ being the Boltzmann constant. 
The position representation of this operator is:

\begin{equation}
\rho(x,x';f) = \langle x | \hat{\rho}(f) | x' \rangle =
 \sum_{n=0}^{\infty} e^{-\beta E_n(f)} \psi_n(x;f) \psi_n^{*}(x';f)  \; .
\label{rhoxxp}
\end{equation}

The unnormalised particle's probability density 
is given by the diagonal elements
of the density matrix, $\rho(x,x;f)$. The canonical partition function
in the presence of the external force 
is defined by the trace of the density matrix:

\begin{equation}
Z(f) = 
       \int_{-\infty}^{\infty} dx \; \rho(x,x;f) = 
       \sum_{n=0}^{\infty} e^{-\beta E_n(f)} \;  ,
\label{z_f}
\end{equation}

where the second equality is obtained by substitution of 
$\rho(x,x;f)$ by the expression given in Eq. (\ref{rhoxxp}).

\subsection{Path integral formulation of the density matrix}
\label{2B}

In the path integral formulation the elements of the unnormalised
density matrix in the presence of the external force are given by:
\cite{feynman65}

\begin{equation}
\rho(x,x';f) = \int_{x\equiv x(0)}^{x' \equiv x(\beta\hbar)} 
D[x(u)] \, e^{-\frac{1}{\hbar}S[x(u);f]} \; .
\label{path_integral}
\end{equation}

$u$ is an imaginary time that 
varies between 0 and $\beta\hbar$, and      
$S[x(u);f]$ is the functional of the Euclidean action 
of the path $x(u)$:

\begin{equation}
S[x(u);f] = \int_{0}^{\beta\hbar} du \, \left(
             \frac{m}{2} \dot{x}(u)^2 + V[x(u)] - fx(u) \right) \; .
\label{action}
\end{equation}

The function $\dot{x}(u)$ represents the derivative of $x(u)$ with
respect to $u$.
The integral measure of the path integral is given by:

\begin{equation}
D[x(u)] = \lim_{N\to\infty} 
          \prod_{k=1}^{N-1} dx_k 
          \left(\frac{mN}{2\pi\beta{\hbar}^2}\right)^{\frac{N}{2}} \; ,
\end{equation}

where the path $x(u)$ has been discretized as 
$(x,x_1,x_2,...,x_{N-1},x')$. \cite{feynman65}

\subsection{Euclidean action under an external force}
\label{2C}

It is important to note the effect of a constant external force on the
Euclidean action of a given path $x(u)$. As the external force
is independent of the imaginary time $u$,
it can be taken out of the time integral in Eq.
(\ref{action}) with the result:

\begin{equation}
S[x(u);f] = S[x(u)] - f\beta\hbar X \; ,
\label{action_property}
\end{equation}

where $S[x(u)]$ is the Euclidean action in the absence of the force,
and $X$ is the average point of the path, or path centroid:

\begin{equation}
X = \frac{1}{\beta\hbar} \int_{0}^{\beta\hbar} du \, x(u) \; .
\end{equation}

For the set of paths contributing to the path integral 
in Eq. (\ref{path_integral}), the relation of having the same
path centroid is an equivalence relation. Then,
the sum over paths may be decomposed into               
a sum over equivalence classes. \cite{cuccoli95} A class
of paths is the subset formed by all paths that have the same centroid
coordinate $X$.
We note that if two paths, say $x(u)$ and
$x'(u)$, belong to the same class,
then the contribution of the external force
to the Euclidean action, 
[i.e., the term $- f\beta\hbar X$ in Eq. (\ref{action_property})],
is identical for both paths. In other words, {\it the contribution of a
constant external force to the Euclidean action
is identical for all paths that 
have the same centroid.}
This simple property is the origin of the interest of 
fixed centroid path integrals in statistical physics.

\subsection{Static-force response density matrix and centroid density}
\label{2D}

The most important quantities to be defined for the class of
paths with centroid at $X$ are the SFR density matrix
\cite{cuccoli95} and the centroid density \cite{cao94}.
We have introduced a new name, {\it static-force response density matrix},
because we will show that
this matrix depends on the response of the system to 
constant external forces of arbitrary magnitude. The name
{\it reduced density matrix} has been used before for this quantity. 
\cite{cuccoli95}
However, this name is less convenient as it is
often encountered in another context.
The SFR density matrix is a constrained path integral
over a given class of paths,
defined by introducing a delta function in the integrand
of Eq. (\ref{path_integral}):

\begin{equation}
\sigma(x,x';X;f) = \int_{x}^{x'} D[x(u)]
   \delta \left(X-\frac{1}{\beta\hbar}\int_{0}^{\beta\hbar}du\,x(u)\right)
   e^{-\frac{1}{\hbar}{S[x(u);f]}}  \; .
\label{sigma_f}
\end{equation}

The centroid density, $C(X;f)$, in the presence of the external force $f$,
is defined as the trace of the SFR density matrix:

\begin{equation}
C(X;f) = \int_{-\infty}^{\infty} dx \; \sigma(x,x;X;f)   \; .
\label{C_f}
\end{equation}

By introducing the expression of the Euclidean action,
[Eq. (\ref{action_property})],
into Eq. (\ref{sigma_f}), we derive an equation that relates
the SFR density matrix in the presence of the external force $f$,
with the same quantity in absence of the force:

\begin{equation}
\sigma(x,x';X;f) = \sigma(x,x';X) \; e^{\beta fX} \; .
\label{sigma_f_property}
\end{equation}

By setting $x=x'$ in the last equation and integrating over $x$, we get 
the following analogous relation for the centroid density:

\begin{equation}
C(X;f) = C(X) \; e^{\beta fX} \; .
\label{C_f_property}
\end{equation}

The last two equations show that if the centroid density,
$C(X)$,  and the SFR
density matrix, $\sigma(x,x';X)$, are known for a given class of paths
in absence of an external force, 
then the corresponding
quantities in the presence of the force $f$ are easily obtained by
multiplication with the constant factor $e^{\beta fX}$.
Dividing the left and right hand sides of the last two equations, we get:

\begin{equation}
[C(X;f)]^{-1} \sigma(x,x';X;f) = [C(X)]^{-1} \sigma(x,x';X)  \, .
\end{equation}

This equation implies that the 
normalized SFR density matrix defined for a given centroid
position $X$ is an invariant 
quantity with respect to the value of the force $f$.
The normalization constant is the trace of the matrix, i.e., the
centroid density.
Setting $x=x'$, we get:

\begin{equation}
\phi^2(x;X) \equiv [C(X;f)]^{-1} \sigma(x,x;X;f) \; .
\label{wave_packets}
\end{equation}

$\phi^2(x;X)$ represents, as a function of $x$,
a normalized particle's probability density.
The normalization 
of this probability density is easily checked by integrating
Eq. (\ref{wave_packets}) with respect to the variable $x$,
and with the help of the definition given in Eq. (\ref{C_f}):              

\begin{equation}
\int_{-\infty}^{\infty} dx \; \phi^2(x;X) = 1             \; .
\end{equation}
This relation is valid irrespective of the value of the 
centroid coordinate, $X$.

The results of this Section may help 
to understand path integral results
that are usually 
presented from a different point of view.
For example, the CMD approximation \cite{cao94} is an 
approximate dynamics for a normalized
density matrix whose diagonal elements in the position
representation are given by $\phi^2(x;X)$. 
An important property of this
density matrix is its invariance
with respect
to the application of an external force to the particle.
Associated to each centroid position, $X$, there is 
a different density matrix,
and the approximate dynamics of CMD 
is formulated through a dynamical equation for the time evolution 
of the centroid coordinate. 

We are now in position to relate the SFR density matrix and
the centroid density to physical observables, an essential step
for the understanding of important path integral concepts
from the point of view of the Schr\"odinger formulation.

\subsection{Moment generating function of the centroid density}
\label{2E}

After the decomposition of the path integral given in
Eq. (\ref{path_integral}) into 
disjoint classes, we can recover
the whole path integral, which is the quantity 
related to physical observables,
by an integral over all classes.
The elements of the unnormalised density matrix, 
in the presence of an external
force $f$, are then obtained as:

\begin{equation}
\rho(x,x';f) = \int_{-\infty}^{\infty} dX \; \sigma(x,x';X;f)  \; .     
\end{equation}

Introducing in this equation the expression
given in Eq. (\ref{sigma_f_property}) for the SFR density matrix,
 one gets:

\begin{equation}
\rho(x,x';f) = \int_{-\infty}^{\infty} dX \; \sigma(x,x';X) \; e^{\beta fX}
\; .
\label{sigma_transform}
\end{equation}

Setting $x=x'$ and integrating over the variable $x$ [with the help of
Eqs. (\ref{z_f}) and (\ref{C_f})], one arrives at:

\begin{equation}
Z(f) = \int_{-\infty}^{\infty} dX \; C(X) \; e^{\beta fX}  \; .
\label{z_transform}
\end{equation}

The last two equations provide the essential link for the correspondence
between fixed centroid path integrals and the Schr\"odinger formulation.
Although 
Eq. (\ref{z_transform}) can be found in Kleinert's book,
\cite{kleinert93} the physical implications of this relation have
not been explored until recently. \cite{ramirez98}
The physical content of Eqs. (\ref{sigma_transform}) and 
(\ref{z_transform}) can be derived by      
two complementary points of view, as suggested by the
mathematical structure of these equations: 
\begin{itemize}
\item
The centroid density, $C(X)$, in absence of the external
force, is related to the partition function, $Z(f)$, 
by an integral
transformation defined by the kernel $e^{\beta fX}$.
The same relation holds between the SFR density matrix,
$\sigma(x,x';X)$, and the density matrix $\rho(x,x';f)$.
This integral transformation
is performed with respect to the 
centroid variable $X$, and the product $\beta f$
is the corresponding conjugate variable. Note that a general property
of an integral transformation is that the physical information contained in
the direct and transformed functions is identical.
Therefore the centroid density $C(X)$, as a function of $X$, 
carries out the same physical information
as the partition function $Z(f)$, as a function of $f$. 
The same relation holds for
$\sigma(x,x';X)$ and $\rho(x,x';f)$. This integral
transformation is called a two-sided Laplace transform, 
with properties similar to those of a Fourier transform. \cite{laplace}

\item
The relation between the centroid density, $C(X)$, and the 
partition function in the presence of a force, $Z(f)$, can be 
interpreted as the relation between a probability density
for the variable $X$ ant its moment generating function. The same
relation holds for $\sigma(x,x';X)$ and $\rho(x,x';f)$. In the
following, we develop this point of view.
\end{itemize}

The moments of the centroid density in the presence of an external force
are defined as:

\begin{equation}
\left\{ X^n\right\}_f  = 
\frac {\int_{-\infty}^{\infty} dX \, C(X;f) \, X^n}
      {\int_{-\infty}^{\infty} dX \; C(X;f)} =  
[Z(f)]^{-1} \int_{-\infty}^{\infty} dX \; C(X;f) \; X^n  \; .
\label{def_momentos}
\end{equation}

The brackets
$\left\{...\right\}_ f$ indicate an average over the 
centroid density, $C(X;f)$, and
the second equality is obtained from Eq. (\ref{z_transform}).
The moments of $X$ in absence of the external force are:

\begin{equation}
\left\{ X^n\right\}  = 
Z^{-1} \int_{-\infty}^{\infty} dX \; C(X) \; X^n  \; ,
\end{equation}

where the partition function is $Z \equiv Z(0)$. Note that the subindex
"$f$" is omitted from the brackets when $f=0$.
The definition of the moment generating function, 
\cite{estadistica} $M(\beta f)$, of the 
normalized centroid density, $Z^{-1}C(X)$, is:

\begin{equation}
M(\beta f) = 
\left\{ e^{\beta fX} \right\} =
Z^{-1} \int_{-\infty}^{\infty} dX \, C(X) \,e^{\beta fX} \; .
\label{moment}
\end{equation}

From Eqs. (\ref{z_transform}) and (\ref{moment}) we get:

\begin{equation}
M(\beta f) = \frac{Z(f)}{Z} \; .
\label{cociente_Z}
\end{equation}

This result has a clear              
physical meaning: {\it the ratio
of partition functions $Z(f)/Z$ is the function generating the moments
of the centroid density}. 
Therefore, these moments 
may be defined as: \cite{estadistica}

\begin{equation}
\left\{ X^n\right\} = 
\left[\frac{\partial M(\beta f)}{\partial(\beta f)}\right]_{f=0} \; ,
\end{equation}

and with the help of Eq. (\ref{cociente_Z}), we get:

\begin{equation}
\left\{ X^n\right\} =  \frac{1}{Z \; {\beta^n}}
           \left[\frac{\partial^n Z(f)}{\partial f^n}\right]_{f=0} \; .
\end{equation}

The l.h.s. of the last equation, i.e., the moments of the
centroid density, is a quantity typically defined
by fixed centroid path integrals, while the r.h.s. contains 
physical quantities, that are defined within the Schr\"odinger 
formulation. 
It is more interesting to study the
cumulant generating function,\cite{estadistica,ma85} $K(\beta f)$,
of the centroid density, which is defined as 
the logarithm of the moment generating function:

\begin{equation}
K(\beta f) = \ln \frac{Z(f)}{Z} = -\beta [F(f)-F] \; ,
\label{k_f}
\end{equation}

where $F(f)$ is the free energy derived from the partition function:
$Z(f) = \exp[-\beta F(f)] $ and $F \equiv F(0)$. 
The cumulants, $\kappa_n$, of the
centroid density are obtained as:

\begin{equation}
\kappa_n = 
\left[\frac{\partial K(\beta f)}{\partial(\beta f)}\right]_{f=0} \; .
\end{equation}

With the help of Eq. (\ref{k_f}) we get:

\begin{equation}
\kappa_n = -\frac{1}{\beta^{n-1}}
      \left[\frac{\partial^nF(f)}{\partial f^n}\right]_{f=0} \; .
\end{equation}

The last equation shows that the cumulants of the
centroid density are related to the
change in the free energy as a constant
external force is acting on the particle.
The first cumulants of the centroid density 
correspond to the mean value and the dispersion of the centroid coordinate:

\begin{equation}
\kappa_1 = \left\{ X\right\} =
  -\left[ \frac{\partial F(f)}{\partial f} \right]_{f=0} \; ,
\label{k1}
\end{equation}

\begin{equation}
\kappa_2 = \delta X^2 =
\left\{ X^2\right\} - \left\{ X\right\}^2 =
  - \frac{1}{\beta}   
  \left[\frac{\partial^2F(f)}{\partial f^2}\right]_{f=0} \; .
\label{k2}
\end{equation}

The centroid dispersion, $\delta X^2$, has been 
related with a "classical delocalization" of the particle.
\cite{gillan90} The precise
physical meaning of this quantity
will be presented in Sec. \ref{3}. 

As the mathematical structure of Eqs. (\ref{sigma_transform})
and (\ref{z_transform}) is identical, the derivation done for 
the moment and cumulant generating functions of the centroid density
can be repeated step by step for the SFR density matrix.
The most important result of this analysis may be
enunciated as follows: {\it
the cumulant generating function, $K_{\sigma}(\beta f)$,
of the SFR density matrix,
$\sigma(x,x';X)$, is the logarithm of the ratio of the elements of
the density matrix in the presence and in absence of the external force
$f$}:

\begin{equation}
K_{\sigma}(\beta f) = \ln\frac{\rho(x,x';f)}{\rho(x,x')} \; .
\label{ln_rho}
\end{equation}

As an application of the analysis presented so far we derive
in Appendix \ref{app2} the centroid density and the SFR density
matrix for a harmonic oscillator by means of their cumulant generating
functions. The information needed for this task is the 
value of the partition function, $Z(f)$, and the 
density matrix, $\rho(x,x';f)$, as a function of the external force $f$.
This example shows how results previously
obtained by solving fixed centroid path integrals \cite{cuccoli95}
can be easily derived in the Schr\"odinger formulation. 

To close this theoretical Section, we explain the precise
meaning of considering the centroid density
as a classical-like  density for the quantum particle, which
is one of the suggestive pictures derived from the 
QTST and CMD approximations. 
Let $\rho^{cla}(x)$ be the classical limit of the 
unnormalised particle's probability density 
in absence of an external force. Then
the classical partition function $Z^{cla}(f)$,
in the presence of a constant external force, is given by:

\begin{equation}
Z^{cla}(f) = \int_{-\infty}^{\infty} dx \, \rho^{cla}(x) \, e^{\beta fx} \; .
\end{equation}

This equation has formally the same structure as Eq. (\ref{z_transform})
for the centroid density, and illustrates in which sense the centroid 
density behaves as a classical density for the quantum particle. In the
classical limit the ratio $Z^{cla}(f)/Z^{cla}$ is the moment generating
function of the classical particle's probability density, while in the
quantum case, the ratio $Z(f)/Z$ is the moment generating
function of the centroid density.

The unnormalised particle's probability 
density is given in the classical case
by a function of the potential energy:

\begin{equation}
\rho^{cla}(x) =
         \left( \frac{m}{2 \pi {\hbar}^2 \beta} \right)^{\frac{1}{2}}
         e^{-\beta V(x)}  \; ,
\end{equation}

In analogy to the classical result, 
the ECP, $F_{ecp}(X)$, for the quantum particle
is defined from the centroid density as:

\begin{equation}
C(X) = \left( \frac{m}{2 \pi {\hbar}^2 \beta} \right)^{\frac{1}{2}}
         \; e^{-\beta F_{ecp}(X)} \; .
\label{ecp}
\end{equation}

We remark that the ECP is sometimes called the
potential of mean force, \cite{gillan90}, the centroid potential, \cite{cao94}
the effective centroid potential, \cite{cao94} 
or the quantum effective potential.
\cite{sese95} However, these names refer all to the same physical quantity.
   
\section{Physical implications}
\label{3}
In this section we focus on some physical implications 
readily derived from the correspondence between fixed centroid path
integrals and the Schr\"odinger formulation. Firstly we present some results 
valid at arbitrary temperatures.  

\subsection{Static susceptibility}
\label{3A}

The first moments of the centroid density have a clear       
physical meaning. We recall that the average position 
of the quantum particle in
thermal equilibrium in the presence of an external force, $f$,
is given by:    

\begin{equation}
\bar{x}(f) = Z(f)^{-1} \int_{-\infty}^{\infty} dx \; x \; \rho(x,x;f) \; .
\end{equation}

From the definition of
the SFR density matrix and of the centroid
density [Eqs. (\ref{sigma_f}) and (\ref{C_f})],
it is easy to show that the average
centroid position, defined from Eq. (\ref{def_momentos}) by
setting $n=1$, and 
the average particle position are identical quantities:

\begin{equation}
\bar{x}(f) = 
\left\{ X\right\}_f = - \frac{\partial F(f)}{\partial f} \; ,
\label{mean_x}
\end{equation}

where the last equality is consequence of an exact generalization
of Eq. (\ref{k1}), derived for $f=0$, to an arbitrary value of the external 
force.
From Eqs. (\ref{k2}) and (\ref{mean_x}),
we obtain the following expression for
the dispersion of the centroid coordinate: 

\begin{equation}
\delta X^2 =
\left\{ X^2\right\} - \left\{ X\right\}^2 =
  \frac{1}{\beta} \left[    
  \frac{\partial \bar{x}(f) }{\partial f} \right]_{f=0} \; .
\label{dispersion}
\end{equation}

The derivative that appears in the last equation is the {\it static
isothermal susceptibility}, $\chi_{xx}^T$, 
a quantity that describes the static response of
the average position of the quantum particle, with respect
to the application of an external force. \cite{kubo,ma85} We have then:

\begin{equation}
\delta X^2 = \frac{1}{\beta} \chi_{xx}^T \; .
\label{susceptibilidad}
\end{equation}

This isothermal susceptibility, $\chi_{xx}^T$, is related to 
{\it the canonical correlation} of the position operator,
$\langle \hat{x};\hat{x} \rangle$ by the following relation: \cite{kubo}

\begin{equation}
\chi_{xx}^T = \beta \left( \langle \hat{x};\hat{x} \rangle -               
                         \bar{x}^2 \right)   \; ,
\label{correlation}
\end{equation}

where $\bar{x} \equiv \bar{x}(0)$ and
the canonical correlation is defined as: \cite{kubo}

\begin{equation}
\langle \hat{x};\hat{x} \rangle = 
 Z^{-1} \beta^{-1}
 \int_{0}^{\beta}d\lambda {\text{Tr}}
 \left[ \exp(-\beta \hat{H})\exp(\lambda\hat{H}) \hat{x}
        \exp( -\lambda \hat{H}) \hat{x} \right] \; .
\end{equation}

Comparing the Eqs. (\ref{susceptibilidad}) and (\ref{correlation}),
and noting that $\bar{x} = \left\{ X \right\} $,
we deduce that the second moment of the centroid density is identical to
the canonical correlation of the position operator:

\begin{equation}
\left\{ X^2 \right\}= \langle \hat{x};\hat{x} \rangle   \; .
\label{xx}
\end{equation}

This result is consistent with the analysis presented
in Ref. \cite{cao94} concerning the relation between the centroid time
correlation function and the Kubo transformed position correlation function.
Eq. (\ref{xx}) is interesting not only to understand the physical
meaning of the centroid density, but also for
practical applications.
In the case of a harmonic
oscillator of angular frequency $\omega$ 
the dispersion of the centroid coordinate,
$\delta X^2$, is given in Eq. (\ref{deslo_clasica}). 
The static isothermal susceptibility for the harmonic case is then: 

\begin{equation}
\chi_{xx}^T = \beta \; \delta X^2 = \frac{1}{m\omega^2} \; .
\label{chixx}
\end{equation}

For an arbitrary anharmonic potential, 
we expect that at low temperatures the static
isothermal susceptibility will be determined by a frequency close to
the first excitation energy, $\Delta E$, of the system.
Then, by substitution in the last equation of
$\omega$ by $\Delta E_{app}/\hbar$, we get the following approximation
to $\Delta E$:

\begin{equation}
\Delta E \approx \Delta E_{app} = 
\hbar \left( \frac{k_BT}{m \delta X^2} \right)^{\frac{1}{2}}
\label{deltae_approx}
\end{equation}

The capability of this approximation has been checked as a 
function of temperature for several one-dimensional 
model potentials, that are listed in
Table \ref{tabla1}. The potentials $V_2$, $V_4$, and
$V_{10}$ are power functions of $x$, whose coefficients
were chosen so that a particle of mass 16 au displays the same 
value of the excitation energy in each of these potentials. 
$V_{dw}$ is a double-well
potential where the first excitation energy corresponds to the
tunnel splitting. By Monte Carlo path integral simulations \cite{ramirez97}
we have obtained the value of the centroid dispersion,
$\delta X^2$, as a function
of temperature, and the excitation
energy has been estimated by Eq. (\ref{deltae_approx}). The results 
are shown
in Fig. \ref{figura1}, where the value of
$\Delta E_{app}$
and the thermal energy, $k_B T$, are  
displayed in units of the $\Delta E$ associated
to each potential. For the harmonic potential, $V_2$,
the approximation is exact at all temperatures. For the potentials
$V_4$ and $V_{10}$ the approximation is remarkably good at temperatures
where the thermal energy is lower than about $1/4$
of the first excitation energy. In this temperature range the
tunnel splitting energy is approximated with an error
of about 25 \%, a value that at least is of 
the correct order of magnitude. This approximation can be applied 
to many-body problems, using the centroid dispersions resulting from the
diagonalization of the tensor given in Eq. (\ref{k2_3d}).

\subsection{$T \to 0$ limit}
\label{3B}

In the zero temperature limit, the ECP 
and the normalized particle's probability 
densities, $\phi^2(x;X)$, [defined in Eq. (\ref{wave_packets})] have 
a simple physical meaning: they are related to the eigenvalues
and eigenfunctions of the Hamiltonian $\hat{H}(f)$.

We are going to combine two pieces of information to
obtain the low temperature limit of $\phi^2(x;X)$. Firstly, 
we know that, as $T \to 0$, the normalized
particle's probability density derived from the path integral
in Eq. (\ref{path_integral}) converges towards the 
probability density of the ground state 
of the Hamiltonian $\hat{H}(f)$. The average value
of the position operator $\hat{x}$ for this ground state is    
$\bar{x}_0(f)$.
Secondly, we have derived that the dispersion of the centroid
coordinate goes to zero in this limit [see Eq. (\ref{dispersion})].
This implies that the centroid density, $C(X;f)$, must be a
delta function of $X$ centered at its average value
$\left\{ X\right\}  = \bar{x}_0(f)$
[see Eq. (\ref{mean_x})].
Therefore only the class of paths with centroid coordinate
at $X=\bar{x}_0(f)$ contributes to the path integral in 
Eq. (\ref{path_integral}). As a consequence, the
normalized probability density, $\phi^2(x;X)$, 
derived for the class of paths with
centroid at $X=\bar{x}_0(f)$ should be identical to the ground state
probability density of the particle 
in the presence of the external force $f$:

\begin{equation}
\lim_{T \to 0} \phi^2(x;X) = |\psi_0(x;f)|^2 \;, 
\text{for} \;\; X \equiv \bar{x}_0(f) \; .
\label{g_limit}
\end{equation}

As illustration of this result we display
in Fig. \ref{figura2} 
the probability densities, $\phi^2(x;X)$, obtained
by fixed centroid Monte Carlo simulations of a particle in the model
potentials $V_2$, $V_4$, and $V_{10}$ at a temperature of 
$k_B T = 0.001$ au, which
is about 1/300 of the first excitation energy, and therefore a good
approximation to the low temperature limit. For each model potential, 
the simulations were performed at three different
centroid positions, $X$. The ground state
densities $|\psi_0(x;f)|^2$ obtained by numerical solution of the
time independent Schr\"odinger equation, for different values of the
external force, $f$, are also shown in the figure. 
Both probability densities are identical, apart from
tiny deviations (not visible in the scale of the figure)
due to the finite temperature and the statistical
uncertainty of the simulation. In the harmonic case the   
displayed probability densities, $\phi^2(x;X)$, differ 
only by a rigid displacement 
as a function of the 
centroid position $X$. We note that these curves are identical to
the probability densities of the coherent states of the
harmonic oscillator. \cite{coherentes} 

In the following we show that
the functions $\phi^2(x;X)$ are related to a variational
principle. Suppose that we look for a quantum state, $| \Psi \rangle$,
of the particle with Hamiltonian $\hat{H}$ whose mean energy, 

\begin{equation}
E_{min}(X) = \langle \Psi | \hat{H} | \Psi \rangle \; ,
\end{equation}

is minimum
against small variations of $|\Psi\rangle$. 
Moreover, this state must satisfy two constraints: its
mean position is fixed at an arbitrary value
$X$, and it is normalized:          

\begin{equation}
X =\langle \Psi | \hat{x} | \Psi \rangle \; ,
\label{xf}
\end{equation}

\begin{equation}
1 = \langle \Psi | \Psi\rangle \; .
\label{1}  
\end{equation}

By straightforward application of calculus of variations (see
Appendix \ref{app3}),
one finds that $ |\Psi\rangle$ must be the ground state of 
a Hamiltonian $\hat{H}(f)$,
i.e., $|\Psi\rangle \equiv |\psi_0(f)\rangle$:

\begin{equation}
(\hat{H} - f \hat{x}) |\psi_0(f)\rangle = E_0(f) |\psi_0(f)\rangle  \, ,
\label{Hf}
\end{equation}

where the force $f$ and the corresponding 
ground state energy, $E_0(f)$, of the
Hamiltonian $\hat{H}(f)$, appear as Lagrange multipliers. The 
value of $f$ must be chosen so that the constraint
in Eq. (\ref{xf}) is satisfied.
The fixed mean position
$X$ corresponds then to the average position of the ground
state $|\psi_0(f)\rangle$, i.e. $X \equiv \bar{x}_0(f)$.
The average energy, $E_{min}(X)$, of these minimum energy states
is derived as a function of $X$ with the help of
Eqs. (\ref{Hf}) and (\ref{xf}) as:

\begin{equation}
E_{min}(X) = E_0(f) + f X \;, \text{for} \;\; X\equiv\bar{x}_0(f)\;  .
\label{Emin}
\end{equation}

We call the states $|\psi_0(f)\rangle$ the
MEWP's of the unperturbed Hamiltonian $\hat{H}$. \cite{ramirez98}
We can now reinterpretate Eq. (\ref{g_limit}) by saying that,
in the zero temperature
limit, the probability density $\phi^2(x;X)$ corresponds to the MEWP's
of the Hamiltonian $\hat{H}$. A characteristic property of these states is
that their average energy is stationary (i.e., minimum)
with respect to any arbitrary change in their probability density
that leaves constant their average position $X$. In the following
we show that this average energy is identical
to the ECP.

As $T \to 0$ we have derived that 
the centroid density, $C(X;f)$, 
is a delta function centered at the average position of the
ground state of $\hat{H}(f)$, i.e. $ \left\{ X\right\}_f  = \bar{x}_0(f)$. 
We also know that 
the integral of $C(X;f)$ with respect to $X$ is identical to the 
partition function $Z(f)$ [see Eqs. (\ref{z_transform}) 
and (\ref{C_f_property})]. Then, in the zero temperature limit 
we can write:

\begin{equation}
\lim_{T \to 0} C(X;f) = Z(f) \delta[X-\bar{x}_0(f)]    \, .
\label{CyZ}
\end{equation}

The asymptotic behavior of $C(X;f)$ as $T \to 0$
is given by an exponential of the ECP
$\exp[-\beta F_{ecp}(X;f)]$, while the asymptotic behavior 
of $Z(f)$ is given by an exponential of the ground state energy
$\exp[-\beta E_0(f)]$. From Eq. (\ref{CyZ}) the asymptotic behavior
of $C(X;f)$ and $Z(f)$ should be the same at $X=\bar{x}_0(f)$, that is:

\begin{equation}
\lim_{T \to 0} F_{ecp}(X;f) = E_0(f)   \;,
\text{for} \;\; X\equiv\bar{x}_0(f)\;  .
\label{FyE}
\end{equation}

From the property given in Eq. (\ref{C_f_property}) for the centroid
density, and from the definition of the ECP in Eq. (\ref{ecp}) it is
easy to derive a relation valid 
at arbitrary temperature:
 
\begin{equation}
F_{ecp}(X;f) = F_{ecp}(X) - fX         \, .
\label{FyF}
\end{equation}

By substitution of the last expression 
into Eq. (\ref{FyE}) we obtain the desired result:
 
\begin{equation}
\lim_{\beta \to \infty} F_{ecp}(X) = E_0(f) + f X = E_{min}(X) \;,
\text{for} \;\; X\equiv\bar{x}(f)\;  ,
\end{equation}

where the last equality corresponds to Eq. (\ref{Emin}). 
Our final conclusion is that, in the limit $T \to 0$,
the value of the ECP, $F_{ecp}(X)$, is equal to
the average energy of that MEWP 
whose average position coincides with the centroid coordinate $X$.
The MEWP's turn out to be the ground states of the Hamiltonian
$\hat{H}(f)$.
The results obtained in this Subsection may be seen as 
a consequence of the
original variational principle of Feynman and Hibbs \cite{feynman65},
when it is applied to the zero temperature limit. However, the
important role of the external force does not appear in the original
path integral formulation of this variational principle.

The CMD approximation is, in the $T \to 0$ limit, an approximate
dynamics based on MEWP's, which is accurate even for highly
anharmonic potentials. \cite{ramirez98} 
Cao and Voth have shown that CMD reproduces correctly 
the classical limit at high temperatures. \cite{cao94} 
The classical limit of our formulation of 
CMD in terms of MEWP's can be derived by the generalization 
of the results at
$T \to 0$ to arbitrary temperatures using the MEWP's 
given by the excited states $|\Psi_n(f) \rangle$ (see Appendix \ref{app3}).
From this generalization we find that
CMD can be formulated {\it for harmonic systems at arbitrary temperatures}
as an exact MEWP's dynamics, even in the classical limit. \cite{ramirez99}
The most
important applications of CMD are related to condensed phase quantum
dynamics of anharmonic systems. \cite{voth96,klein98,martyna96}
From our analysis in terms of MEWP's we find that
for anharmonic potentials CMD provides accurate results
in both the $T \to 0$ and the classical limits, but further work
is needed to clarify the capability of CMD at intermediate temperatures.

\section{Summary}
\label{4}

We have presented the correspondence between
fixed centroid path integrals and the Schr\"odinger
formulation.
This analysis shows that the Feynman path centroid density 
and the SFR density matrix
depend on the static response of the quantum system
to a constant external force. The 
path centroid density is related by a simple integral transformation
to the partition function of the
quantum system under the action of an external force.
The same integral transformation
is found between a classical phase space
density and the classical partition function.
Therefore, 
the interpretation 
of the centroid density as a classical-like density for the quantum
system, which is one of the ideas suggested by the 
QTST and CMD approximations, has a precise
physical meaning: the path centroid density behaves
"classically" in the sense that under the action of an
external force it has the same static response properties
as a classical phase space density.
The same integral transformation is found between
the normalized canonical density matrix in the presence of an
external force  and the SFR density matrix.
Within the Schr\"odinger formulation it is essential to
introduce a constant external force to define the centroid density 
and the elements of the SFR density matrix as transformed functions. 
However, in the path integral formulation the introduction of
external forces is not needed,
because one works
directly with the transformed functions. 
This facility of the path integral formulation is the origin 
a large number
of path integral applications, from 
a variational approximation of the thermodynamic
properties of quantum system, \cite{feynman65} to the CMD approximation of
time correlation functions of quantum particles
in thermal equilibrium. \cite{cao94} 

In the present work, we have not tried to present the implications
of our formulation in current
applications based on fixed centroid path integrals.
Nevertheless, our analysis has led to results that clarify
the physical meaning of fixed centroid quantities. In particular,
we have shown that the dispersion of the centroid coordinate is
related to the static isothermal susceptibility, and 
that, in the zero temperature limit, the fixed centroid path integrals
are related to the MEWP's of the unperturbed system.
At $T=0$, the normalized SFR density matrix is
a pure state density matrix corresponding to a MEWP of the Hamiltonian,
and the Feynman effective classical potential is
the average energy of the MEWP's.

\acknowledgments
This work was supported by DGICYT (Spain) under contract
PB96-0874. We thank E. Artacho for helpful
discussions, and L.M. Ses\'e and M.C. B\"ohm for critically 
reading the manuscript.

\appendix
\section{Extension to many-particles}
\label{app1}

The results presented for a particle 
moving in one dimension can
be easily generalized to multidimensional $n$-body systems
of distinguishable particles.
We denote the set of $n$ particle coordinates
as a 3$n$-dimensional vector 
$\bf{r} = ( \bf{r_1}, \bf{r_2}, ..., \bf{r_n} )$ where
the $i$-particle position vector is 
${\bf r_i} = (r_{ix}, r_{iy}, r_{iz})$.
The set of external forces is 
$\bf{f} = ( \bf{f_1}, \bf{f_2}, ..., \bf{f_n} )$.
The external force $\bf{f_i}$ acts on the particle $\bf{r_i}$
through the linear term, $-\bf{f_i} \hat{\bf{r}}_i$, appearing in
the potential energy of the Hamiltonian $\hat{H}(\bf{f})$. 
The centroid 
density associated to the $n$-body Hamiltonian 
in absence of external forces, 
$\hat{H}$,
can be defined by a generalization of Eq. (\ref{z_transform}) as:

\begin{equation}
Z(\bf{f}) = \int_{-\infty}^{\infty} ... \int_{-\infty}^{\infty}
 d\bf{r} \; 
C(\bf{X}) \; e^{\beta \bf{f} \bf{X}}  \; ,
\label{z_transform_3d}
\end{equation}
where 
$\bf{X} = ( \bf{X_1}, \bf{X_2}, ..., \bf{X_n} )$
is a 3$n$-dimensional vector formed by the centroid positions
of each particle. The last equation represents a 3$n$-dimensional
two-sided Laplace transform. The cumulants of the centroid density are 
obtained as the derivatives of the free energy $F(\bf{f})$ 
associated to the partition function $Z(\bf{f})$:

\begin{equation}
\left\{ X_{ix} \right\} =
-\left[ \frac{\partial F(\bf{f})}{\partial f_{ix}} \right]_{\bf{f}=\bf{0}} \; ,
\label{k1_3d}
\end{equation}

\begin{equation}
\left\{ X_{ix} X_{jy} \right\} - 
\left\{ X_{ix} \right\}
\left\{ X_{jy} \right\} =
  - \frac{1}{\beta}
\left[\frac{\partial^2F(\bf{f})}
  {{\partial f_{ix}} {\partial f_{jy}} }\right]_{\bf{f}=\bf{0}} \; .
\label{k2_3d}
\end{equation}

The last expression represents the components of a tensor that is related
to the static isothermal susceptibity tensor by multiplication
by the constant $\beta$. 

\section{Centroid density and static-force response density matrix 
for a linear harmonic oscillator} 
\label{app2}

We consider a canonical ensemble of 
particles of mass $m$ moving in a one-dimensional 
harmonic potential $V_{ho}(x)= (1/2)m\omega^2x^2$. In absence of an
external force, the partition function and the free energy are:
\cite{feynman72}

\begin{equation}
Z = \left[2\sinh\left(\frac{\beta\hbar\omega}{2}\right)\right]^{-1} \; ,
\end{equation}

\begin{equation}
F = k_BT\ln\left[2\sinh\left(\frac{\beta\hbar\omega}{2}\right)\right] \; .
\end{equation}

By application of a constant external force, $f$, the potential changes to
$V_{ho}(x)-fx$. The new potential energy is also a quadratic function
of $x$, and the corresponding free energy is easily derived as:         

\begin{equation}
F(f) = F - \frac{f^2}{2m\omega^2}    \; .                              
\end{equation}

From Eqs. (\ref{k1}) and (\ref{k2}) we obtain
the first two cumulants of the centroid density $C(X)$ as: 

\begin{equation}
\left\{ X\right\}  = \left[ \frac{f}{m\omega^2} \right]_{f=0} = 0 \; ,
\end{equation}

\begin{equation}
\delta X^2 = 
 \frac{1}{\beta m\omega^2}  \; .
\label{deslo_clasica}
\end{equation}

The higher order cumulants, which are proportional 
to successive derivatives of the last equation with respect to $f$,
are zero. This result implies that the centroid density must be a
Gaussian function of $X$:

\begin{equation}
C(X) = N_1 \; G_X(\bar{X};\delta X^2)   \; ,
\end{equation}

where $N_1$ is a normalization constant and
$G_X(\bar{X};\delta X^2)$ is a
normalized Gaussian function of $X$:

\begin{equation}
G_X(\bar{X};\delta X^2) = \left( \frac{1}{2\pi\delta X^2} \right)^{\frac{1}{2}} 
    \exp\left[-\frac{(\bar{X}-X)^2}{2\delta X^2}\right] \; .
\end{equation}

The constant $N_1$ 
can be obtained from Eq. (\ref{z_transform})
by setting $f=0$:

\begin{equation}
Z = \int_{-\infty}^{\infty} dX C(X) = N_1
\label{Z_def}
\end{equation}

The centroid density of the harmonic oscillator is then:

\begin{equation}
C(X) = Z \; G_X\left(0;\frac{1}{\beta m \omega^2}\right) \; .
\label{C_def}
\end{equation}

The last result has been derived several times using 
fixed centroid path integrals.
\cite{cuccoli95,gillan90,kleinert93,cao90}
The ECP for the harmonic oscillator is obtained
from its definition in Eq. (\ref{ecp}) and the 
last equation as: 

\begin{equation}
F_{ecp}(X) = F - k_BT\ln\left(\beta\hbar\omega\right)
             + \frac{1}{2}m\omega^2X^2  \; .
\end{equation}

The SFR density matrix $\sigma(x,x';X)$ may be derived following
the same scheme. The cumulant generating function in
this case was given in Eq. (\ref{ln_rho}).
The elements of the density matrix for the harmonic oscillator 
in absence of an external force are:
\cite{feynman72}

\begin{equation}
\rho(x,x') = 
\left[ \frac{m\omega}{2\pi\hbar\sinh(\beta\hbar\omega)} \right]^{\frac{1}{2}}
       \exp\left\{ -\frac{m\omega}{2\hbar\sinh(\beta\hbar\omega)}
       \left[ (x^2+x'^2)\cosh(\beta\hbar\omega)-2xx'\right] \right\} \; .
\end{equation}

While in the presence of an external force we obtain from the
definition in Eq. (\ref{rhoxxp}):

\begin{equation}
\rho(x,x';f) = \exp\left(\frac{\beta f^2}{2m\omega^2}\right) 
\rho(x-x_{min},x'-x_{min}) \; ,
\end{equation}

where we have used the following relations:

\begin{equation}
E_n(f) = E_n - \frac{f^2}{2m\omega^2}  \; ,   
\end{equation}
\begin{equation}
\psi_n(x;f) = \psi_n(x-x_{min})  \; .   
\end{equation}

$x_{min}=f/(m\omega^2)$ is the position of minimum energy for
the potential $V_{ho}(x)-fx$. The cumulants of the 
centroid variable $X$, with $\sigma(x,x';X)$ as its probability
density, are evaluated as the derivatives of the cumulant generating
function [Eq. (\ref{ln_rho})] with respect to the force at $f=0$. We find
that only the first and second cumulants $(\kappa_1, \kappa_2)$ 
are different from zero.
Therefore $\sigma(x,x';X)$ must be a Gaussian function of the
variable X, with $\kappa_1$ being the mean value of $X$, and
$\kappa_2$ its dispersion:
 
\begin{equation}
\sigma(x,x';X) = N_2 G_X(\kappa_1;\kappa_2)   \; .
\end{equation}

The result for the first cumulant is obtained from the cumulant
generating function
after straightforward algebra as:

\begin{equation}
\kappa_1=\frac{\delta x^2_{cla}}{\delta x^2}\left(\frac{x+x'}{2}\right) \; ,
\end{equation}

and the result for the second cumulant is:

\begin{equation}
\kappa_2=\frac{\delta x^2_{cla}}{\delta x^2}
         (\delta x^2-\delta x^2_{cla}) \; .
\end{equation}

The constants $\delta x^2$ and $\delta x^2_{cla}$ 
correspond to the dispersion
of the position coordinate for a canonical ensemble of harmonic oscillators
in the quantum and classical case,
respectively:

\begin{equation}
\delta x^2 = \frac{\hbar}{2m\omega}
  \coth\left(\frac{\beta\hbar\omega}{2}\right) \; ,
\end{equation}
\begin{equation}
\delta x^2_{cla} = \frac{1}{\beta m\omega^2} \; .
\end{equation}

The normalization constant, $N_2$, is determined from 
Eq. (\ref{sigma_transform}) by setting $f=0$. We get:  

\begin{equation}
N_2 = \rho(x,x') \; .
\end{equation}

The final result is:

\begin{equation}
\sigma(x,x';X) = \rho(x,x') 
  G_X\left[ \frac{\delta x^2_{cla}}{\delta x^2}\left(\frac{x'+x}{2}\right);
         \frac{\delta x^2_{cla}}{\delta x^2}
         (\delta x^2-\delta x^2_{cla})  \right]  \; .
\end{equation}

The elements of the SFR density matrix are the product of the 
elements of the canonical density matrix by a normalized Gaussian
function of the centroid coordinate $X$. 
This equation is identical 
to Eq. (B.4) of Ref. \cite{cuccoli95} (apart from a different
grouping of factors).

\section{Minimum energy wave packets}
\label{app3}

We want to find the MEWP's associated to the Hamiltonian 
$\hat{H}$.
To simplify the notation we use the following abbreviations: 
$\Psi \equiv \langle x | \Psi \rangle $, 
$\Psi' \equiv \partial\Psi/\partial x$,
$\Psi'' \equiv \partial^2\Psi/\partial x^2$, $V \equiv V(x)$,
and $D \equiv -{\hbar}^2/(2m)$.

The MEWP's minimize the functional:           

\begin{equation}
E_{min} = \int_{-\infty}^{\infty} dx \; (D\Psi\Psi''+V\Psi^2)
\label{functional}
\end{equation}

with the constraints given in Eqs. (\ref{xf},\ref{1}) for the wave function
$\Psi$.
We call:

\begin{equation}
A = D\Psi\Psi''+V\Psi^2 - f x \Psi^2  -E \Psi^2 ,
\end{equation}

where $f$ and $E$ are two Lagrange multipliers. 
From the Euler equation: 

\begin{equation}
\frac{\partial A}{\partial\Psi} -
\frac{d}{dx}\frac{\partial A}{\partial\Psi'} +
\frac{d^2}{dx^2}\frac{\partial A}{\partial\Psi''} = 0  \; ,
\end{equation}

one derives the differential equation that must
be satisfied by the MEWP's:

\begin{equation}
D\Psi'' + (V-fx) \Psi = E\Psi  \; ,
\end{equation}

which is the time independent Hamilton equation 
corresponding to the Hamiltonian 
$\hat{H}(f)$. All functions satisfying this equation,
i.e., the eigenfunction $|\Psi_n(f) \rangle$, are MEWP's in the sense
that they minimize the functional in Eq. (\ref{functional}) subject to the 
constrains given in Eqs. (\ref{xf},\ref{1}). The excited states
($n > 1$) correspond to local minima of the functional.
In the study of the $T \to 0$ limit of the SFR density matrix we use
only the states $|\Psi_0(f) \rangle$ corresponding to the
global minimum of the functional, but the excited states are 
necessary to generalize this study to arbitrary temperatures.

\begin{figure}
\caption{
First excitation energy estimated from the
dispersion of the centroid coordinate for four model potentials
at different temperatures. For anharmonic potentials the 
approximation works only at low temperature.
For each model potential the true excitation energy, $\Delta E$, 
was used as energy unit.
}
\label{figura1}
\end{figure}

\begin{figure}
\caption{
Normalized probability densities,
$\phi^2(x;X)$,
obtained by fixed centroid path integral simulations
at $k_BT$ = 0.001 au,
using the model potentials $V_2$, $V_4$, and $V_{10}$. 
For each potential model,
the centroid coordinate was fixed at $X = -0.5$, 0, and 0.5 au,
respectively. Superposed to the simulation results there are
the probability
densities of the ground states of these potentials in the presence
of an external force $f$. The value of $f$ associated to each
ground state is given in au.
}
\label{figura2}
\end{figure}

\narrowtext
\begin{table}
\caption{
One-dimensional model potentials. 
The lowest energy eigenvalues ($E_0$, $E_1$, and $E_2$) are given
for a particle of mass 16 au. The first
excitation energy is $\Delta E = E_1 - E_0$. The values were
obtained by numerical solution of the time independent 
Schr\"odinger equation. All units in au.
}
~\newline
\begin{tabular}{lcccc}
potential    & $E_0$ & $E_1$ & $E_2$ & $\Delta E$ \\
\tableline
V$_2$ = $x^2$               &  0.177 & 0.530 & 0.884 & 0.354 \\
V$_4$ = 2.2015 $x^4$        &  0.137 & 0.490 & 0.962 & 0.354 \\
V$_{10}$ = 21.797 $x^{10}$  &  0.121 & 0.474 & 1.038 & 0.354 \\
V$_{dw}$ = 0.25 $(x^2-1)^2$ &  0.143 & 0.172 & 0.371 & 0.029 \\
\end{tabular}
\label{tabla1}
\end{table}

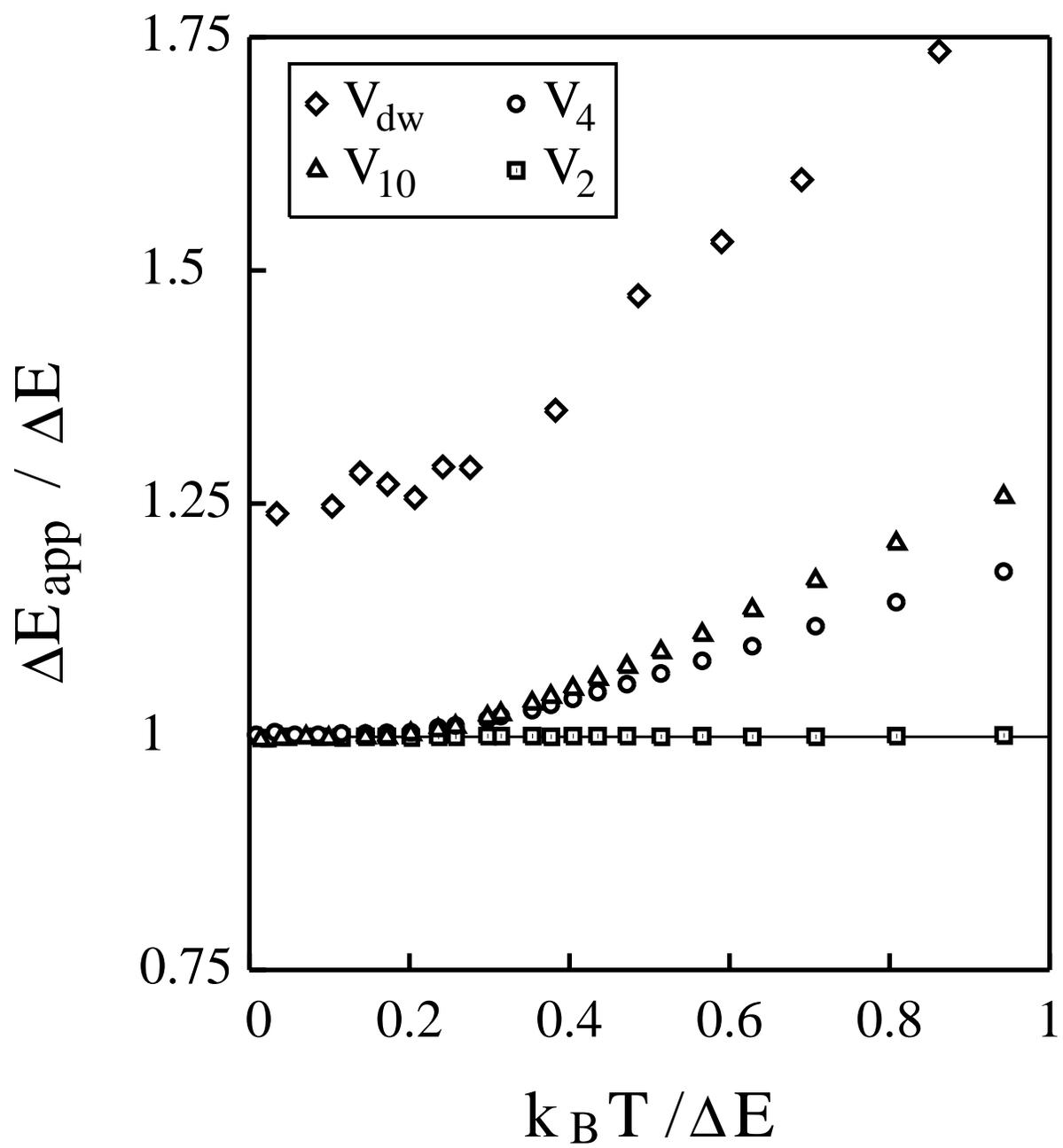
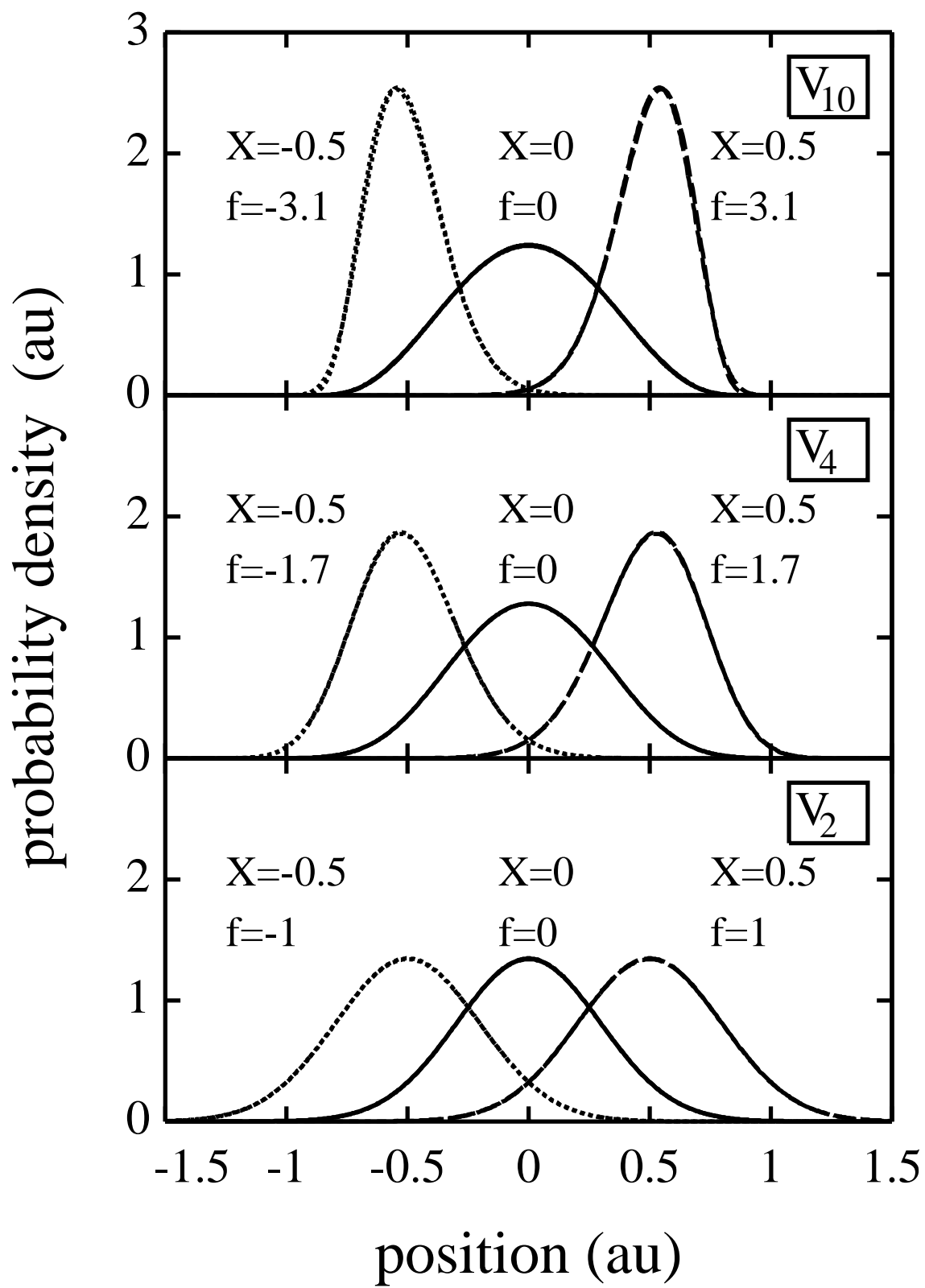
\end{document}